\documentclass[aps,prd,amsmath,amssymb,superscriptaddress,preprintnumbers,twocolumn,10pt,nofootinbib]{revtex4-1}
\usepackage{graphicx}
\usepackage{dcolumn}
\usepackage{bm}
\usepackage{amssymb}
\usepackage{latexsym}
\usepackage{booktabs}
\usepackage{amsmath}
\usepackage{multirow}
\usepackage{url}
\usepackage{footnote}
\usepackage{float}
\usepackage{threeparttable}
\usepackage[colorlinks=true, linkcolor=blue, citecolor=blue]{hyperref}
\usepackage[bottom]{footmisc}

\usepackage[normalem]{ulem}
\usepackage{color}
\usepackage{array}
\usepackage{enumerate}
\usepackage{adjustbox}

\usepackage{makecell}
\usepackage{diagbox}
\usepackage{epstopdf}
\usepackage{epsfig}
\usepackage{longtable}
\usepackage{supertabular}
\usepackage{algorithm}
\usepackage{pifont}
\usepackage{algorithmic}
\usepackage{changepage}
\usepackage{setspace}
\IfFileExists{orcidlink.sty}{\usepackage{orcidlink}}{%
  \newcommand{\orcidlink}[1]{%
    \href{https://orcid.org/##1}{%
      \includegraphics[height=1.6ex]{orcid-ID.png}%
    }%
  }%
}

\begin{document}

\title{Updated constraints on interacting dark energy: A comprehensive analysis using multiple CMB probes, DESI DR2, and supernovae observations}

\author{Tian-Nuo Li}
\affiliation{Liaoning Key Laboratory of Cosmology and Astrophysics, College of Sciences, Northeastern University, Shenyang 110819, China}

\author{Guo-Hong Du}
\affiliation{Liaoning Key Laboratory of Cosmology and Astrophysics, College of Sciences, Northeastern University, Shenyang 110819, China}

\author{Yun-He Li}
\affiliation{Liaoning Key Laboratory of Cosmology and Astrophysics, College of Sciences, Northeastern University, Shenyang 110819, China}

\author{Yichao Li}
\affiliation{Liaoning Key Laboratory of Cosmology and Astrophysics, College of Sciences, Northeastern University, Shenyang 110819, China}

\author{Jia-Le Ling}
\affiliation{Institute of Theoretical Physics, Chinese Academy of Sciences, Beijing 100190, China}

\author{Jing-Fei Zhang}
\affiliation{Liaoning Key Laboratory of Cosmology and Astrophysics, College of Sciences, Northeastern University, Shenyang 110819, China}

\author{Xin Zhang}\thanks{Corresponding author}\email{zhangxin@neu.edu.cn}
\affiliation{Liaoning Key Laboratory of Cosmology and Astrophysics, College of Sciences, Northeastern University, Shenyang 110819, China}
\affiliation{MOE Key Laboratory of Data Analytics and Optimization for Smart Industry, Northeastern University, Shenyang 110819, China}
\affiliation{National Frontiers Science Center for Industrial Intelligence and Systems Optimization, Northeastern University, Shenyang 110819, China}

\begin{abstract}
Recent DESI baryon acoustic oscillation (BAO) measurements, combined with Planck cosmic microwave background (CMB) data and DESY5 type Ia supernova (SN) data, indicate a significant deviation from $\Lambda$CDM, which seems to suggest that this deviation can be explained by an interaction between dark energy and dark matter. In this work, we perform a comprehensive analysis by utilizing the latest DESI DR2 BAO data in conjunction with CMB data from ACT, SPT, Planck, and WMAP, along with SN data from PantheonPlus and DESY5. We consider four interacting dark energy (IDE) models with different forms of the interaction term $Q$. Our analysis indicates that CMB experiments other than Planck enhance the evidence for an interaction in the IDE models with $Q \propto \rho_{\rm de}$. In particular, when using the SPT+DESI+DESY5 data, the IDE model with $Q = \beta H_0 \rho_{\rm de}$ gives $\beta = -0.4170 \pm 0.1220$, with a deviation from zero reaching $3.4\sigma$ level. When replacing DESY5 with PantheonPlus, this deviation weakens to $2.1\sigma$ level, but remains relatively significant. Furthermore, the Bayes factors of the IDE model with $Q = \beta H_0 \rho_{\rm de}$ are positive in all cases, providing a moderate-to-strong preference over $\Lambda$CDM. Overall, our comprehensive analysis clearly suggests that the IDE models with $Q \propto \rho_{\rm de}$ (especially, $Q = \beta H_0 \rho_{\rm de}$) provide strong evidence supporting the existence of interaction and are more preferred by the current cosmological data.
\end{abstract}

\keywords{interacting dark energy, cosmological parameters, cosmological observations, observational constraints}

\maketitle

\section{Introduction}
 
Observational evidence from the type Ia supernova (SN) \cite{SupernovaSearchTeam:1998fmf,SupernovaCosmologyProject:1998vns}, large-scale structures \cite{SDSS:2005xqv}, and cosmic microwave background (CMB) \cite{WMAP:2003elm} has robustly confirmed that the universe is undergoing a phase of late-time accelerated expansion. The accelerated expansion of the universe is commonly attributed to dark energy (DE), which is interpreted as a negative-pressure fluid, with the simplest candidate being Einstein's cosmological constant, $\Lambda$, characterized by an equation of state (EoS) parameter $w = -1$. The standard cosmological model, known as $\Lambda$CDM, incorporates the cosmological constant $\Lambda$ and cold dark matter (CDM), which has been remarkably successful in explaining most cosmological and astrophysical observations.

However, advances in cosmological observations have revealed certain tensions within the $\Lambda$CDM framework, such as the $S_8$ tension \cite{DiValentino:2020vvd} and the $H_0$ tension \cite{Verde:2019ivm}. Specifically, the most significant case is the $H_0$ tension, with a discrepancy exceeding $5\sigma$ between the results inferred from the Planck CMB observations under the $\Lambda$CDM assumption \cite{Planck:2018vyg} and the SH0ES's local measurements based on the Cepheid-calibrated distance ladder \cite{Riess:2021jrx}. In recent years, the $H_0$ tension has been a central topic in current cosmology (see, e.g., refs.~\cite{Zhao:2019gyk,Bernal:2016gxb,Guo:2018ans,Vagnozzi:2019ezj,Cai:2021wgv,Vagnozzi:2021gjh,Vagnozzi:2021tjv,Yang:2021eud,Escudero:2022rbq,James:2022dcx,Song:2022siz,Jin:2022qnj,Pierra:2023deu,Vagnozzi:2023nrq,Huang:2024erq,Breuval:2024lsv,Zhang:2024rra,Song:2025ddm,Jin:2025dvf}; for relevant reviews, see also, e.g., refs.~\cite{DiValentino:2021izs,Kamionkowski:2022pkx}). Additionally, the $\Lambda$CDM model suffers from two major problems related to the cosmological constant, namely the ``fine-tuning'' and ``cosmic coincidence'' problems~\cite{Sahni:1999gb,Bean:2005ru}. These issues challenge the completeness of the $\Lambda$CDM model and motivate the search for alternative cosmological scenarios \cite{Boisseau:2000pr,Chevallier:2000qy,Li:2004rb,Zhang:2005hs,Zhang:2007sh,Ma:2007av,Zhang:2007bi,Zhang:2009un,Cui:2009ns,Fu:2011ab,Zhang:2014ifa,Cai:2015emx,Nojiri:2017ncd,Feng:2016djj,Zhang:2015rha,Guo:2015gpa,Zhang:2014nta,Feng:2017usu,Feng:2017nss,Feng:2019jqa,Poulin:2018cxd,Yin:2023srb,Yao:2023qve,Yao:2023ybs}; for a recent comprehensive review, see ref. \cite{CosmoVerseNetwork:2025alb}.

Among the various extended scenarios, one notable class is the interacting dark energy (IDE) models, which explicitly consider direct non-gravitational interactions between DE and dark matter (DM). In such an IDE scenario, the overall evolution and expansion history of the universe, as well as the growth of matter and structure formation, will differ from the $\Lambda$CDM model. It has been found that the IDE models can not only help alleviate the  cosmic coincidence problem \cite{Chimento:2003iea,Hu:2006ar,Dutta:2017kch}, $S_8$ tension \cite{Lucca:2021dxo}, and $H_0$ tension \cite{DiValentino:2017iww,Yang:2018euj,DiValentino:2019ffd,Gao:2021xnk}, but also help to probe the fundamental nature of DE and DM. In recent years, the IDE models have been extensively discussed in the literature \cite{Farrar:2003uw,Zhang:2004gc,Cai:2004dk,Zhang:2005rg,Zhang:2005rj,Wang:2006qw,Li:2011ga,Zhang:2012uu,Li:2014cee,Li:2015vla,Costa:2016tpb,Pan:2019gop,Li:2019ajo,Yao:2020pji,Pan:2020zza,Yao:2020hkw,Yang:2021oxc,Wang:2021kxc,Zhang:2021yof,Yao:2022kub,Li:2023gtu,Giare:2024ytc,vanderWesthuizen:2025vcb,vanderWesthuizen:2025rip,vanderWesthuizen:2025mnw,Wang:2005ph,He:2010im,Wang:2007ak} (see refs.~\cite{Wang:2016lxa,Wang:2024vmw} for relevant reviews). In particular, our previous work used the baryon acoustic oscillation (BAO) data from the first data release (DR1) of the Dark Energy Spectroscopic Instrument (DESI), combined with the Planck CMB and DESY5 SN data, to constrain the IDE models, and found evidence for existence of an interaction at the $3\sigma$ confidence level \cite{Li:2024qso}.

Recently, the DESI has publicly released its second data release (DR2), which includes BAO measurements from more than 14 million extragalactic sources over the redshift range $0.1 \leq z \leq 4.2$, thereby providing unprecedented constraints on cosmic distance measurements. When combined with DESI DR2 BAO data, the SN datasets (PantheonPlus, Union3, and DESY5), together with CMB data from Planck and Atacama Cosmology Telescope (ACT), reveal a $2.8\sigma$--$4.2\sigma$ preference for dynamical DE\footnote{The EoS parameter of DE evolves across $w = -1$ (transitioning from a past phantom-like behavior to a present quintessence-like behavior), a phenomenon known as the Quintom scenario \cite{Feng:2004ad,Guo:2004fq,Zhang:2005yz,Zhang:2005kj,Zhang:2006qu}.} within the $w_0w_a$CDM model \cite{DESI:2025zgx}. These significant deviations from the cosmological constant, as highlighted by DESI, have sparked extensive debates on the nature of DE \cite{Giare:2024gpk,Dinda:2024ktd,Escamilla:2024ahl,Sabogal:2024yha,Li:2024qus,Wang:2024dka,DESI:2024kob,DESI:2025fii,Huang:2025som,Li:2025owk,Wu:2025wyk,Li:2025ula,Li:2025ops,Barua:2025ypw,Yashiki:2025loj,Ling:2025lmw,Goswami:2025uih,Yang:2025boq,Pang:2025lvh,You:2025uon,Ozulker:2025ehg,Cheng:2025lod,Pan:2025qwy,Fazzari:2025lzd,Park:2025fbl,Alam:2025epg,Toomey:2025xyo,Hussain:2025uye,Goh:2025upc,Wolf:2025acj,Yadav:2025vpx,Yang:2025uyv,Wang:2025znm,Shah:2025ayl,Silva:2025hxw,Luciano:2025ykr,Samanta:2025oqz,Plaza:2025nip,Guedezounme:2025wav,Moretti:2025gbp,Silva:2025twg,Qiang:2025cxp,Li:2025vuh,Li:2026xaz,Li:2026asg}, as well as on various other aspects of cosmological physics (see, e.g., refs.~\cite{RoyChoudhury:2024wri,Yang:2025ume,Chen:2025wwn,Wang:2025zri,Kumar:2025etf,Abedin:2025dis,Khoury:2025txd,Araya:2025rqz,Li:2025eqh,Li:2025dwz,Paliathanasis:2025xxm,Du:2024pai,Jiang:2024viw,Du:2025iow,Feng:2025mlo,RoyChoudhury:2025dhe,Du:2025xes,Zhou:2025nkb,Cai:2025mas,Li:2025cxn,Odintsov:2025jfq,Nojiri:2025low,Avila:2025sjz,Li:2025htp,Jia:2025poj,RoyChoudhury:2025iis,Pedrotti:2025ccw,Paliathanasis:2025kmg,Wang:2025vtw,Liu:2025myr,Yao:2025kuz,Adam:2025kve,Lopez-Hernandez:2025lbj,Yang:2025oax,Dinda:2025hiu,Camarena:2025upt,Gomez-Valent:2025mfl,Afroz:2025iwo,Yang:2025qdg,Liu:2025evk,Wang:2025gus,Liu:2025mob,Paul:2025wix,Braglia:2025gdo,Guin:2025xki,Du:2026cly}). 

Given the importance of the result, its robustness has been thoroughly tested in the literature \cite{Giare:2024gpk,Colgain:2024xqj,Huang:2024qno,Wang:2024pui}. Giarè \emph{et al.} \cite{Giare:2024gpk} noted that when combining CMB, DESI BAO, and SN data, the preference for dynamical DE evidence varies depending on the choice of different parameterizations. Colgáin \emph{et al.} \cite{Colgain:2024xqj} investigated potential systematic effects in DESI BAO measurements that may have contributed to the reported preference for dynamical DE, with the DESI BAO measurement at $z = 0.51$ showing approximately a $2\sigma$ tension with predictions based on the Planck best-fit $\Lambda$CDM cosmology. Additionally, the impact of SN measurements was recently highlighted in ref.~\cite{Efstathiou:2024xcq}, where cross-correlating the PantheonPlus and DESY5 SN data revealed a calibration difference of approximately 0.04 mag between low and high redshifts\footnote{Discussions on the consistency of SN measurements across high and low redshifts can be found, see, e.g., refs.~\cite{Colgain:2022nlb,Colgain:2022rxy,Malekjani:2023ple}, as well as other discussions surrounding potential calibration issues in DESY5, see also ref.~\cite{Colgain:2024ksa}.}. Some studies suggest that the evidence for dynamical DE may primarily originate from systematics in the DESY5 SN data, as the parameter range favored by the uncorrected DESY5 sample diverges from many other cosmological datasets \cite{Notari:2024zmi,Huang:2024qno}.

In addition, in recent months, both the ACT \cite{AtacamaCosmologyTelescope:2025blo} and the South Pole Telescope (SPT) \cite{SPT-3G:2025bzu}, ground-based CMB experiments, have released new measurements of temperature and polarization anisotropies. Although these measurements partially overlap with Planck data, they extend into higher multipole regions (i.e., smaller angular scales) and represent the most precise measurements of small-scale CMB polarization to date. When considered independently, and especially in combination with the large-scale CMB data from WMAP 9-year observations \cite{WMAP:2012nax}, they provide significant new constraints on cosmological parameters. Recently, some cosmological studies utilizing ACT and SPT data have been conducted \cite{Peng:2025tqt, Peng:2025vda, Pang:2024wul, Giare:2024oil, AtacamaCosmologyTelescope:2025nti, Poulin:2025nfb, SPT-3G:2025vyw}. For example, Peng and Piao \cite{Peng:2025vda} constrained the periodic oscillations in the primordial power spectrum using the latest ACT and SPT CMB data, combined with Planck CMB data, providing state-of-the-art CMB constraints on primordial oscillations. Giarè \cite{Giare:2024oil} tested the robustness of the preference for dynamical DE using different CMB data (ACT, SPT, and their combination with WMAP or Planck data), along with DESI BAO and SN data, and found that CMB experiments other than Planck generally weaken the evidence for dynamical DE.

Driven by these motivations and concerns, it is essential and important to conduct a comprehensive analysis of IDE models, utilizing the latest cosmological observational data. In this work, we use the latest CMB data from ACT, SPT, and their combinations with WMAP or Planck, along with DESI DR2 and SN data from PantheonPlus and DESY5, to constrain IDE models. To achieve high generality, we investigate four different forms of the interaction term $Q$. In addition, we utilize Bayesian evidence to assess which form of the IDE model is preferred by the current observational data. We will present a comprehensive and robust result regarding whether the current observational data support the interaction between DE and DM, and which types of IDE models are more consistent with the current cosmological observations.

This work is organized as follows. In Section~\ref{sec2}, we briefly introduce the models considered in this work, along with the cosmological data utilized in the analysis. In Section~\ref{sec3}, we report the constraint results and make some relevant discussions. The conclusion is given in Section~\ref{sec4}.

\section{Methodology and data}\label{sec2}

\subsection{Brief description of the IDE models}\label{sec2.1}

For the IDE model, the modified conservation equations for the energy-momentum tensor ($T_{\mu\nu}$) of DE and CDM are written as 
\begin{equation}
  \label{eq:energyexchange} \nabla_\nu T^\nu_{\hphantom{j}\mu,\rm de} = -\nabla_\nu T^\nu_{\hphantom{j}\mu,\rm c}  = Q_\mu,
\end{equation} 
where $Q_\mu$ represents the energy-momentum transfer vector.

At the background level, the continuity (energy conservation) equations can be written as
\begin{align}\label{conservation1}
{\rho}_{\rm de}^{\prime } +3\mathcal{H}(1+w)\rho_{\rm de}= aQ,\\
{\rho}_{\rm c}^{\prime } +3 \mathcal{H} \rho_{\rm c}= -aQ,
\end{align}
where prime denotes differentiation with respect to the conformal time, $\rho_{\rm{de}}$ and $\rho_{\rm c}$ represent the energy densities of DE and CDM, $w$ is the EoS parameter of DE. Here, $\mathcal{H}= a H$ is the conformal Hubble parameter, and $Q$ denotes the interaction term describing the energy transfer rate between DE and CDM.

For the interaction term $Q$, in the absence of a fundamental theory, we adopt a phenomenological approach in which $Q$ is assumed to be proportional to the energy densities of either DE or CDM \cite{Amendola:1999qq, Billyard:2000bh}. To ensure consistency in dimensions, it is multiplied by a factor that has units of inverse time. The most natural choice for this is the Hubble parameter, as it provides an analytical solution to the conservation equations. Thus, the interaction between DE and CDM can be represented phenomenologically as $Q = \beta H \rho_{\rm de}$ or $Q = \beta H \rho_{\rm c}$, where $\beta$ is the dimensionless coupling parameter. $\beta > 0$ indicates CDM decaying into DE, $\beta < 0$ indicates DE decaying into CDM, and $\beta = 0$ indicates no interaction between DE and CDM. In addition, some researchers in the field of IDE argue that $Q$ should not depend on the Hubble parameter, as local interactions should not rely on the global expansion of the universe \cite{Valiviita:2008iv,Boehmer:2008av,Caldera-Cabral:2008yyo,He:2008si,Clemson:2011an}. Following this viewpoint, another form of $Q$ is assumed, such as $Q = \beta H_0 \rho_{\rm c}$ or $Q = \beta H_0 \rho_{\rm de}$, where the appearance of $H_0$ is solely for dimensional reasons. It should be emphasized that we only wish to extend the base $\Lambda$CDM model in a minimal way. Thus, we focus solely on the case where $w = -1$ to avoid introducing additional parameters, as also adopted in related studies \cite{Xu:2011qv,Chimento:2013rya,Wang:2014xca,SanchezG:2014snn,Guo:2017hea,Wang:2021kxc}. We refer to this IDE model with $w = -1$ as I$\Lambda$CDM, where the four typical phenomenological forms of $Q$ are given by: $Q = \beta H \rho_{\rm de}$ (I$\Lambda$CDM1), $Q = \beta H \rho_{\rm c}$ (I$\Lambda$CDM2), $Q = \beta H_0 \rho_{\rm de}$ (I$\Lambda$CDM3), and $Q = \beta H_0 \rho_{\rm c}$ (I$\Lambda$CDM4).
 
At the linear perturbation level, eq.~(\ref{eq:energyexchange}) is given by
\begin{widetext}
\begin{equation}
 {\delta\rho_I'} + 3\mathcal{H}({\delta \rho_I}+ {\delta p_I})+(\rho_I+p_I)(k{v}_I + 3 H_L') = a(\delta Q_I + AQ_I), \label{eqn:conservation1}
\end{equation}
\begin{equation}
[(\rho_I + p_I)(v_I - B)]' + 4\mathcal{H}(\rho_I + p_I)(v_I - B) - k \delta p_I + \frac{2}{3}k c_K p_I \Pi_I - k (\rho_I + p_I) A = a[Q_I(v - B) + f_I], \label{eqn:conservation2}
\end{equation}
\end{widetext}
where $I$ represents $\rm DE$ or $\rm CDM$, $\delta\rho_I$ is energy density perturbation, $\delta p_I$ is isotropic pressure perturbation, $\delta Q_I$ and $f_I$ are the energy and momentum transfer rate perturbation, respectively. Here, $v$ denotes the total velocity perturbation of all fluid components. $A$, $B$, and $H_L$ represent the scalar metric perturbations, $v_I$ is velocity perturbation, $\Pi_I$ is anisotropic stress perturbation, and $c_K = 1-3K/k^2$ with $K$ being the spatial curvature. To streamline our notation, we introduce 
\begin{equation}
  f_{k,I}={f_I\over k},\quad\theta_I={v_I-B\over k}. \label{eq:fk_theta}
\end{equation}
In our work, $Q_{\rm de}=-Q_{\rm c}=Q$, $\delta Q_{\rm de}=-\delta Q_{\rm c}=\delta Q$, and $f_{\rm de}=-f_{\rm c}=f$, suggesting that when $Q$, $\delta Q$, and $f$ are all positive, both energy and momentum transfer from CDM into DE.

Building on our previous work \cite{Li:2023fdk}, we directly parameterize the energy transfer perturbation and the momentum transfer potential as follows:

\begin{align}
  \delta Q & = C_1 \delta_{\rm de} + C_2 \delta_{\rm c} + C_3 \theta_{\rm de}, \label{eq:parametrized_dQ} \\
  f_k      & = -Q \theta + D_1 \theta_{\rm de} + D_2 \theta_{\rm c}, \label{eq:parametrized_f}
\end{align}
where $\delta_{\rm de}$($\delta_{\rm c}$) and $\theta_{\rm de}$($\theta_{\rm c}$) represent the fractional density and velocity divergence perturbations for DE (CDM), respectively. The unsubscripted quantity $\theta$ denotes the total velocity divergence perturbation of all fluid components. The coefficients $ C_1 $, $ C_2 $, $ C_3 $, $ D_1 $, and $ D_2 $ are time-dependent functions. Different IDE models correspond to distinct functional forms for these coefficients. As an example, for the $ Q = \beta H \rho_{\rm de} $ model considered in this study, the corresponding coefficient functions are $ C_1 = D_2 = Q $ and $ C_2 = C_3 = D_1 = 0 $. Based on this parameterization, we only need to solve for the model-specific functions in each IDE scenario to determine the perturbation evolution of DE and CDM. For further details, refer to ref.~\cite{Li:2023fdk}.

When computing the perturbation evolution of DE in IDE models, it is important to note that vacuum energy ($w = -1$) does not constitute a true geometric background. According to standard linear perturbation theory, DE is treated as a non-adiabatic fluid with negative pressure. The interaction between DE and DM affects the non-adiabatic pressure perturbations of DE and may lead to the divergence of non-adiabatic curvature perturbations on large scales, resulting in the so-called large-scale instability problem \cite{Majerotto:2009zz,Clemson:2011an}. In other words, cosmological perturbations of DE within IDE models may diverge in certain regions of parameter space, which could cause the breakdown of IDE cosmology at the perturbation level. To address this issue, the parameterized post-Friedmann (PPF) framework \cite{Fang:2008sn,Hu:2008zd} has been extended to IDE models \cite{Li:2014eha,Li:2014cee,Li:2015vla,Li:2023fdk}, known as the ePPF approach. This methodology enables reliable calculation of cosmological perturbations across the entire parameter space of IDE models. In this work, we apply the ePPF approach to handle cosmological perturbations (see, e.g., refs.~\cite{Zhang:2017ize,Feng:2018yew}, for applications of the ePPF approach).

\subsection{Cosmological data}\label{sec2.2}

In the MCMC analysis, we assume a spatially flat universe, with $K=0$ or equivalently $\Omega_{\rm k}=0$. Table~\ref{tab1} lists the free parameters of these models and the uniform priors applied. The parameter set for the $\Lambda$CDM model is $\bm{\theta}_{\Lambda\mathrm{CDM}}=\{\Omega_{\rm b} h^2$, $\Omega_{\rm c} h^2$, $H_0$, $\tau$, $\log(10^{10}A_{\rm s})$, $n_{\rm s}\}$. For the I$\Lambda$CDM models, the parameter sets are $\bm{\theta}_{\mathrm{I}\Lambda\mathrm{CDM}}=\{\bm{\theta}_{\Lambda\mathrm{CDM}},\beta\}$. Note that we treat $H_0$ as a free parameter instead of the commonly used $\theta_{\rm {MC}}$, as it depends on a standard non-interacting background evolution. The theoretical models are computed using a modified version of the {\tt CAMB} code \cite{Lewis:1999bs}, which allows for interactions between DE and DM\footnote{\url{https://github.com/liaocrane/IDECAMB}.} \cite{Li:2023fdk}. We use the publicly available sampler {\tt Cobaya} \cite{Torrado:2020dgo} to perform Markov Chain Monte Carlo (MCMC) analysis, and assess the convergence of the MCMC chains using the Gelman-Rubin statistic, with the condition $R - 1 < 0.02$ \cite{Gelman:1992zz}. The MCMC chains are analyzed using the public package {\tt GetDist} \cite{Lewis:2019xzd}. We use the latest observational data to constrain these models and obtain the best-fit values and the $1$--$2\sigma$ confidence level ranges for the interesting parameters of \{$H_{0}$, $\Omega_{\mathrm{m}}$, $\beta$\}. 

\begin{table}[H]
\caption{Flat priors on the main cosmological parameters constrained in this paper.}
\begin{center}
\renewcommand{\arraystretch}{1.8}
\begin{tabular}{@{\hspace{0.4cm}}c@{\hspace{0.6cm}} c@{\hspace{0.6cm}} c @{\hspace{0.4cm}} }
\hline\hline
\textbf{Model}       & \textbf{Parameter}       & \textbf{Prior}\\
\hline
$\Lambda$CDM        & $\Omega_{\rm b} h^2$                     & $\mathcal{U}$[0.005\,,\,0.1] \\
                    & $\Omega_{\rm c} h^2$                     & $\mathcal{U}$[0.01\,,\,0.99] \\
                    & $H_0$                                    & $\mathcal{U}$[20\,,\,100] \\
                    & $\tau$                                   & $\mathcal{U}$[0.01\,,\,0.8] \\
                    & $\log(10^{10}A_{\rm s})$                 & $\mathcal{U}$[1.61\,,\,3.91] \\
                    & $n_{\rm s}$                              & $\mathcal{U}$[0.8\,,\,1.2] \\
\hline
I$\Lambda$CDM       & $\beta$                                  & $\mathcal{U}$[-1\,,\,1] \\

\hline\hline
\end{tabular}
\label{tab1}
\end{center}	
\end{table}

The datasets used are as follows:
\begin{itemize}
    
\item \textbf{\texttt{Planck}:} The Planck CMB likelihoods include the \texttt{Commander} likelihood for the TT spectrum and the \texttt{SimAll} likelihood for the EE spectrum~\cite{Planck:2018vyg,Planck:2019nip}, both in the range $2 \leq \ell \leq 30$; the NPIPE high-$\ell$ \texttt{CamSpec} likelihood for the TT spectrum in the range $30 \leq \ell \leq 2500$, and for the TE and EE spectra in the range $30 \leq \ell \leq 2000$~\cite{Efstathiou:2019mdh,Rosenberg:2022sdy}; as well as the CMB lensing likelihood, utilizing the latest high-precision reconstruction from NPIPE PR4 Planck data\footnote{\url{https://github.com/carronj/planck_PR4_lensing}.}~\cite{Carron:2022eyg}.

\item \textbf{\texttt{ACT}:} The ACT Data Release 6 (DR6), with data collected from 2017 until the conclusion of the experiment in 2022, provides measurements of CMB temperature, polarization, and lensing anisotropies. Specifically, we employ the ACT DR6 likelihood\footnote{\url{https://github.com/ACTCollaboration/act_dr6_mflike}.} (the spectrum in the multipole range $600\leq \ell \leq 8500$)~\cite{AtacamaCosmologyTelescope:2025blo} and the ACT DR6 lensing likelihood\footnote{\url{https://github.com/ACTCollaboration/act_dr6_lenslike}.}~\cite{ACT:2023dou,ACT:2023kun}, together with a Gaussian prior on $\tau=0.0566\pm0.0058$.

\item \textbf{\texttt{SPT}:} The SPT-3G D1 release\footnote{\url{https://github.com/SouthPoleTelescope/spt_candl_data}.}
 \cite{SPT-3G:2025bzu} provides the CMB temperature and polarization (TT, TE, EE) anisotropy spectra from observations of the Main field taken during 2019 and 2020, along with a Gaussian prior on $\tau=0.0566\pm0.0058$.

\item \textbf{\texttt{WMAP}:} The WMAP 9-year observational data release\footnote{\url{https://github.com/HTJense/pyWMAP}.} \cite{WMAP:2012nax} is used, including CMB temperature and polarization data, with low-$\ell$ TE data (which may be contaminated by dust) excluded. The minimum multipole in the TE spectrum is set at $\ell = 24$. When combining WMAP with ACT or SPT data, a Gaussian prior of $\tau=0.0566\pm0.0058$ is always applied.

\item \textbf{\texttt{DESI}:} The DESI DR2 BAO measurements include the transverse comoving distance $D_{\mathrm{M}}/r_{\mathrm{d}}$, the angle-averaged distance $D_{\mathrm{V}}/r_{\mathrm{d}}$, and the Hubble horizon $D_{\mathrm{H}}/r_{\mathrm{d}}$, where $r_{\mathrm{d}}$ denotes the comoving sound horizon at the drag epoch. The measurements utilized in this work are detailed in Table IV of ref.~\cite{DESI:2025zgx}. These measurements are derived from multiple tracers, including the bright galaxy sample, luminous red galaxies, emission line galaxies, quasars, and the Lyman-$\alpha$ forest.

\item \textbf{\texttt{DESY5}:} The DESY5 sample comprises 1829 type Ia supernovae (SNe), assembled from two parts: 1635 photometrically classified SNe from the released portion of the full 5-year data of the Dark Energy Survey collaboration, with redshifts in the range $0.1 < z < 1.3$, and 194 low-redshift SNe from the CfA3~\cite{Hicken:2009df}, CfA4~\cite{Hicken:2012zr}, CSP~\cite{Krisciunas:2017yoe}, and Foundation~\cite{Foley:2017zdq} samples, with redshifts in the range $0.025 < z < 0.1$ \cite{DES:2024jxu}.

\item \textbf{\texttt{PantheonPlus}:} The PantheonPlus comprises 1550 spectroscopically confirmed SNe from 18 different surveys, spanning the redshift range $0.01 < z < 2.26$ \cite{Brout:2022vxf}.

\end{itemize}

\section{Results and discussions}\label{sec3}

\begin{table*}[htbp]
\centering
\caption{Fitting results ($1\sigma$ confidence level) in the $\Lambda$CDM, I$\Lambda$CDM1, I$\Lambda$CDM2, I$\Lambda$CDM3, and I$\Lambda$CDM4 models from the Planck+DESI+DESY5, ACT+DESI+DESY5, SPT+DESI+DESY5, Planck+DESI+PantheonPlus, ACT+DESI+PantheonPlus, and SPT+DESI+PantheonPlus data. Here, $H_{0}$ is in units of ${\rm km}~{\rm s}^{-1}~{\rm Mpc}^{-1}$.}
\label{tab2}
\setlength{\tabcolsep}{6mm}
\renewcommand{\arraystretch}{1.5}
\small
\begin{tabular}{lc c c c}
\hline 
\hline
Model/Dataset & $H_0$ &$\Omega_{\mathrm{m}}$& $\beta$ \\
\hline
$\bm{\Lambda}$\textbf{CDM} &  &  &  \\
Planck+DESI+DESY5 & $68.00\pm0.27$ & $0.3049\pm 0.0035$ & ---  \\
ACT+DESI+DESY5 & $68.23\pm 0.28$ & $0.3037\pm 0.0038$ & ---  \\
SPT+DESI+DESY5 & $68.22\pm 0.31$ & $0.3011\pm 0.0041$ & --- \\
Planck+DESI+PantheonPlus & $68.10\pm0.27$ & $0.3035\pm 0.0035$ & ---  \\
ACT+DESI+PantheonPlus & $68.35\pm 0.28$ & $0.3021\pm 0.0037$ & ---  \\
SPT+DESI+PantheonPlus & $68.35\pm 0.32$ & $0.2993\pm 0.0042$ & --- \\
\hline
$\bm{\textbf{I}\Lambda\textbf{CDM1}}$ &  &  &  \\
Planck+DESI+DESY5 & $67.20\pm 0.53$ & $0.3360\pm 0.0180$ & $-0.1110\pm 0.0640$  \\
ACT+DESI+DESY5 & $67.33\pm 0.54$ & $0.3410\pm 0.0190$ & $-0.1310\pm 0.0670$  \\
SPT+DESI+DESY5 & $66.74\pm 0.58$ & $0.3620\pm 0.0210$ & $-0.2310\pm 0.0760$ \\
Planck+DESI+PantheonPlus & $67.81\pm0.57$ & $0.3105\pm 0.0190$ & $-0.0390\pm 0.0670$  \\
ACT+DESI+PantheonPlus & $67.97\pm 0.56$ & $0.3170\pm 0.0210$ & $-0.0520\pm 0.0710$  \\
SPT+DESI+PantheonPlus & $67.38\pm 0.62$ & $0.3390\pm 0.0220$ & $-0.1460\pm 0.0820$ \\
\hline
$\bm{\textbf{I}\Lambda\textbf{CDM2}}$ &  &  &  \\
Planck+DESI+DESY5 & $68.42\pm0.38$ & $0.3003\pm 0.0045$ & $0.0015\pm 0.0009$  \\
ACT+DESI+DESY5 & $68.95\pm 0.43$ & $0.2980\pm 0.0045$ & $0.0029\pm 0.0013$  \\
SPT+DESI+DESY5 & $68.49\pm 0.59$ & $0.2996\pm 0.0049$ & $0.0013\pm 0.0026$ \\
Planck+DESI+PantheonPlus & $68.59\pm0.37$ & $0.2982\pm 0.0044$ & $0.0017\pm 0.0009$  \\
ACT+DESI+PantheonPlus & $69.14\pm 0.43$ & $0.2956\pm 0.0045$ & $0.0032\pm 0.0013$  \\
SPT+DESI+PantheonPlus & $68.72\pm 0.59$ & $0.2973\pm 0.0050$ & $0.0020\pm 0.0026$ \\
\hline
$\bm{\textbf{I}\Lambda\textbf{CDM3}}$ &  &  &   \\
Planck+DESI+DESY5 & $66.90\pm0.56$ & $0.3580\pm 0.0240$ & $-0.2730\pm 0.1050$  \\
ACT+DESI+DESY5 & $67.00\pm 0.59$ & $0.3640\pm 0.0250$ & $-0.2930\pm 0.1210$  \\
SPT+DESI+DESY5 & $66.58\pm 0.57$ & $0.3840\pm 0.0250$ & $-0.4170\pm 0.1220$ \\
Planck+DESI+PantheonPlus & $67.61\pm0.60$ & $0.3270\pm 0.0260$ & $-0.1220\pm 0.1140$  \\
ACT+DESI+PantheonPlus & $67.69\pm 0.60$ & $0.3330\pm 0.0250$ & $-0.1560\pm 0.1210$  \\
SPT+DESI+PantheonPlus & $67.28\pm 0.60$ & $0.3530\pm 0.0260$ & $-0.2770\pm 0.1330$ \\

\hline
$\bm{\textbf{I}\Lambda\textbf{CDM4}}$ &  &  &    \\
Planck+DESI+DESY5 & $67.85\pm 0.54$ & $0.3090\pm 0.0110$ & $-0.0110\pm 0.0320$  \\
ACT+DESI+DESY5 & $67.85\pm 0.60$ & $0.3130\pm 0.0130$ & $-0.0290\pm 0.0410$  \\
SPT+DESI+DESY5 & $66.97\pm 0.64$ & $0.3320\pm 0.0150$ & $-0.1080\pm 0.0490$ \\
Planck+DESI+PantheonPlus & $68.35\pm0.54$ & $0.2980\pm 0.0110$ & $0.0170\pm 0.0320$  \\
ACT+DESI+PantheonPlus & $68.47\pm 0.61$ & $0.3010\pm 0.0130$ & $0.0096\pm 0.0410$  \\
SPT+DESI+PantheonPlus & $67.75\pm 0.67$ & $0.3160\pm 0.0150$ & $-0.0580\pm 0.0510$ \\

\hline
\hline
\end{tabular}
\end{table*}

In this section, we present the main results of the cosmological analysis for four IDE models: I$\Lambda$CDM1 ($Q=\beta H\rho_{\rm de}$), I$\Lambda$CDM2 ($Q=\beta H\rho_{\rm c}$), I$\Lambda$CDM3 ($Q=\beta H_0\rho_{\rm de}$), and I$\Lambda$CDM4 ($Q=\beta H_0\rho_{\rm c}$). In Section~\ref{sec3.1}, we show the constraints from different combinations of independent CMB data combined with DESI BAO and SN (DESY5, PantheonPlus) data. The main aim of this subsection is to investigate whether CMB experiments other than Planck enhance or diminish the evidence for existence of interaction. In Section~\ref{sec3.2}, we present the results obtained from ACT and SPT small-scale data combined with WMAP or Planck temperature and polarization measurements at large angular scales, as well as DESI BAO and SN (DESY5, PantheonPlus) data. The main aim of this subsection is to explore how the combination of different CMB data affects the evidence for existence of interaction. In Section~\ref{sec3.3}, we apply the Bayesian evidence to assess which IDE model is favored by the current data over $\Lambda$CDM.

\begin{figure*}[htbp]
\includegraphics[scale=0.7]{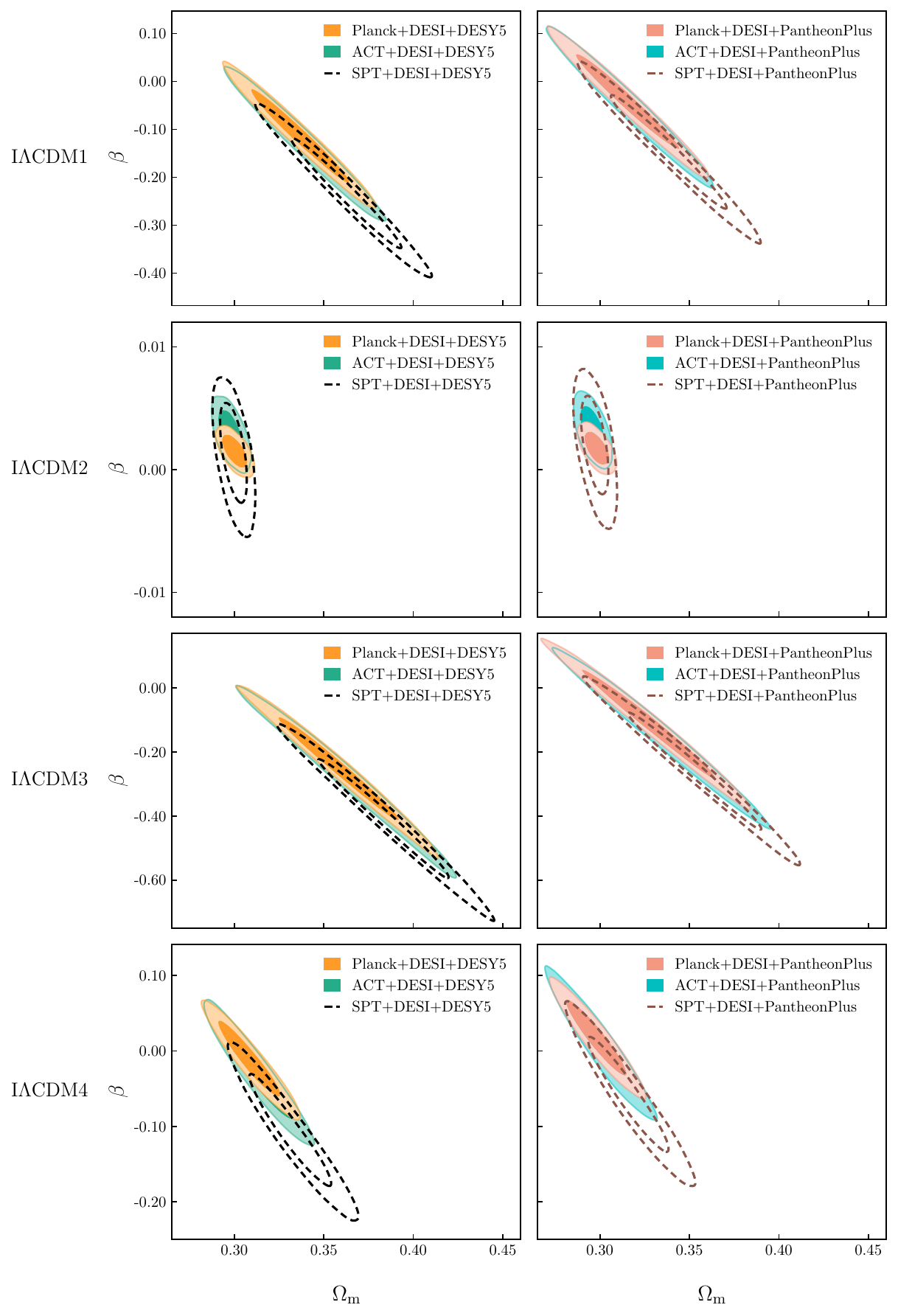}
\centering
\caption{\label{fig1} Two-dimensional marginalized contours ($1\sigma$ and $2\sigma$ confidence levels) in the $\Omega_{\mathrm{m}}$--$\beta$ plane by using the Planck+DESI+DESY5, ACT+DESI+DESY5, SPT+DESI+DESY5, Planck+DESI+PantheonPlus, ACT+DESI+PantheonPlus, and SPT+DESI+PantheonPlus data in the I$\Lambda$CDM1, I$\Lambda$CDM2, I$\Lambda$CDM3, and I$\Lambda$CDM4 models.}
\end{figure*}

\subsection{Constraints from different CMB experiments}\label{sec3.1}

The constraints on the cosmological parameters using data from different independent CMB experiments are summarized in Table~\ref{tab2}, and the two-dimensional probability contours in the $\Omega_{\mathrm{m}}$--$\beta$ plane are shown in Figure~\ref{fig1}.

We begin by reporting the results based on the Planck CMB constraints. In Figure~\ref{fig1}, constraints in the $\Omega_{\mathrm{m}}$--$\beta$ plane are displayed, with the left panel including the Planck+DESI+DESY5 data (orange) and the right panel including those from Planck+DESI+PantheonPlus (light orange). When using the Planck+DESI+DESY5, the constraint values of $\beta$ are $-0.1110\pm 0.0640$ (I$\Lambda$CDM1), $0.0015\pm 0.0009$ (I$\Lambda$CDM2), $-0.2730\pm 0.1050$ (I$\Lambda$CDM3), and $-0.0110\pm 0.0320$ (I$\Lambda$CDM4). Unsurprisingly, the new constraints on the I$\Lambda$CDM model from DESI DR2 in combination with Planck and DESY5 are broadly consistent with our earlier results based on DESI DR1 \cite{Li:2024qso}, with only slight shifts in the central values and a modest improvement in precision. We observe that I$\Lambda$CDM2 yields a result that is different from those of the other I$\Lambda$CDM models, with a much smaller value of the coupling parameter $\beta$. This difference can be understood from the specific form of the interaction term in I$\Lambda$CDM2, namely $Q=\beta H\rho_c$. In the early universe, both the Hubble parameter $H$ and the CDM density $\rho_c$ have relatively large values. As a result, $\beta$ in the I$\Lambda$CDM2 model is required to be very small and can be tightly constrained. Since CMB data provide a powerful probe of the early universe, they are particularly sensitive to this interaction form, leading to the relatively different constraint result obtained for I$\Lambda$CDM2. Therefore, in I$\Lambda$CDM2 the departure of $\beta$ from zero reaches $1.7\sigma$; however, its central value is so small that, even if a non-zero interaction exists, it would be extremely weak and the model effectively reverts to $\Lambda$CDM. For I$\Lambda$CDM4, the deviation of $\beta$ from zero is only $0.3\sigma$, indicating essentially no interaction. By contrast, in I$\Lambda$CDM1 and I$\Lambda$CDM3, the deviations of $\beta$ from zero reach $1.7\sigma$ and $2.6\sigma$, respectively, providing evidence for existence of an interaction. However, several prior studies have found that replacing DESY5 with PantheonPlus weakens the evidence for dynamical DE \cite{DESI:2025zgx,Huang:2024qno}, a phenomenon also reflected in other independent analyses \cite{Giare:2025pzu,Li:2025eqh}. Therefore, we assess whether the evidence for existence of an interaction likewise diminishes when DESY5 is replaced by PantheonPlus. We find that replacing DESY5 with PantheonPlus likewise weakens the indication of a nonzero interaction. For example, the I$\Lambda$CDM1 and I$\Lambda$CDM3 cases yield $\beta = -0.0390 \pm 0.0670,$ and $\beta = -0.1220 \pm 0.1140$, corresponding to deviations from zero of only $0.6\sigma$ and $1.1\sigma$, respectively.

As the next step, we investigate the results based on the ACT CMB constraints. In Figure~\ref{fig1}, constraints in the $\Omega_{\mathrm{m}}$--$\beta$ plane are displayed, with the left panel including the ACT+DESI+DESY5 data (green) and the right panel including those from ACT+DESI+PantheonPlus (light green). When using the ACT+DESI+DESY5, the constraint values of $\beta$ are $-0.1310\pm 0.0670$ (I$\Lambda$CDM1), $0.0029\pm 0.0013$ (I$\Lambda$CDM2), $-0.2930\pm 0.1210$ (I$\Lambda$CDM3), and $-0.0290\pm 0.0410$ (I$\Lambda$CDM4). We find that, relative to Planck+DESI+DESY5 data, ACT+DESI+DESY5 data yield slightly weaker constraints on the coupling parameter $\beta$, with the central value exhibiting a somewhat larger deviation from zero. Therefore, the significance of the deviations of $\beta$ from zero for the IDE models is essentially unchanged, with I$\Lambda$CDM1 slightly increasing to $2.0\sigma$ and I$\Lambda$CDM3 slightly decreasing at $2.4\sigma$. Furthermore, we observe that, upon replacing DESY5 with PantheonPlus, the evidence for existence of an interaction remains similarly diminished. In particular, for I$\Lambda$CDM3, the value of $\beta$ is $ -0.2930\pm0.1210$, indicating the deviation decreases to $1.3\sigma$ compared with ACT+DESI+DESY5.

\begin{figure*}[htpb]
\includegraphics[scale=0.6]{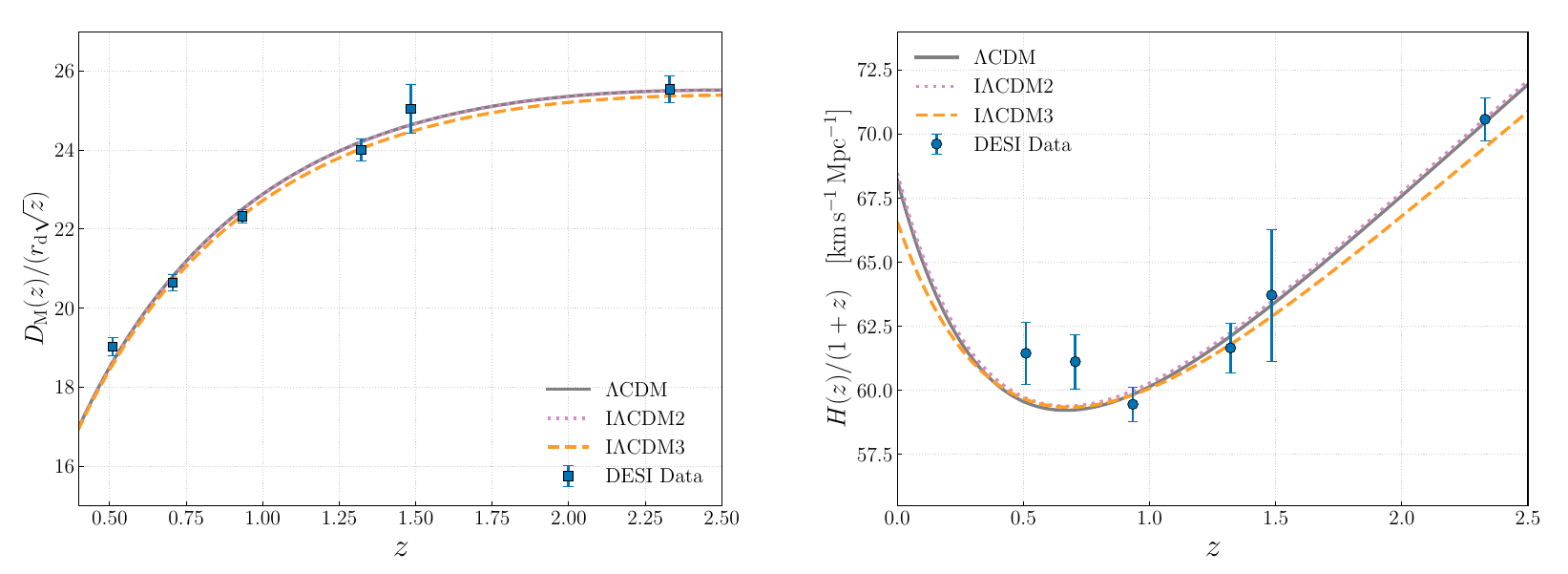}
\centering \caption{\label{fig2} {Best-fit predictions for the rescaled transverse comoving distance $D_\mathrm{M}(z)$ (left panel) and Hubble expansion rate $H(z)$ (right panel) for the $\Lambda$CDM, I$\Lambda$CDM2, and I$\Lambda$CDM3 models obtained from the analysis of SPT+DESI+DESY5 data.}}
\end{figure*}

\begin{table*}[htbp]
\centering
\caption{Fitting results ($1\sigma$ confidence level) in the $\Lambda$CDM, I$\Lambda$CDM1, I$\Lambda$CDM2, I$\Lambda$CDM3, and I$\Lambda$CDM4 models from the ACT+WMAP+DESI+DESY5, SPT+WMAP+DESI+DESY5, ACT+Planck-cut+DESI+DESY5, and SPT+Planck-cut+DESI+DESY5 data. Here, $H_{0}$ is in units of ${\rm km}~{\rm s}^{-1}~{\rm Mpc}^{-1}$.}
\label{tab3}
\setlength{\tabcolsep}{6mm}
\renewcommand{\arraystretch}{1.4}
\small
\begin{tabular}{lc c c c}
\hline 
\hline
Model/Dataset & $H_0$ &$\Omega_{\mathrm{m}}$& $\beta$ \\
\hline
$\bm{\Lambda}$\textbf{CDM} &  &  &  \\
ACT+WMAP+DESI+DESY5 & $68.26\pm 0.26$ & $0.3036\pm 0.0036$ & ---  \\
SPT+WMAP+DESI+DESY5 & $68.26\pm 0.31$ & $0.3011\pm 0.0041$ & ---  \\
ACT+Planck-cut+DESI+DESY5 & $68.23\pm 0.25$ & $0.3025\pm 0.0032$ & ---  \\
SPT+Planck-cut+DESI+DESY5 & $67.98\pm 0.28$ & $0.3037\pm 0.0036$ & ---  \\
\hline
$\bm{\textbf{I}\Lambda\textbf{CDM1}}$ &  &  &  \\
ACT+WMAP+DESI+DESY5 & $67.36\pm 0.55$ & $0.3380\pm 0.0190$ & $-0.1230\pm 0.0670$  \\
SPT+WMAP+DESI+DESY5 & $66.84\pm 0.58$ & $0.3590\pm 0.0210$ & $-0.2190\pm 0.0770$  \\
ACT+Planck-cut+DESI+DESY5 & $67.27\pm 0.55$ & $0.3380\pm 0.0180$ & $-0.1220\pm 0.0630$  \\
SPT+Planck-cut+DESI+DESY5 & $67.09\pm 0.55$ & $0.3380\pm 0.0190$ & $-0.1230\pm 0.0670$  \\
\hline
$\bm{\textbf{I}\Lambda\textbf{CDM2}}$ &  &  &  \\
ACT+WMAP+DESI+DESY5 & $68.75\pm 0.39$ & $0.2992\pm 0.0044$ & $0.0017\pm 0.0011$  \\
SPT+WMAP+DESI+DESY5 & $68.49\pm 0.42$ & $0.2996\pm 0.0045$ & $0.0013\pm 0.0017$  \\
ACT+Planck-cut+DESI+DESY5 & $68.37\pm 0.37$ & $0.3011\pm 0.0043$ & $0.0004\pm 0.0008$  \\
SPT+Planck-cut+DESI+DESY5 & $68.33\pm 0.37$ & $0.3001\pm 0.0044$ & $0.0013\pm 0.0009$  \\
\hline
$\bm{\textbf{I}\Lambda\textbf{CDM3}}$ &  &  &   \\
ACT+WMAP+DESI+DESY5 & $67.04\pm 0.55$ & $0.3620\pm 0.0230$ & $-0.2730\pm 0.1120$  \\
SPT+WMAP+DESI+DESY5 & $66.61\pm 0.57$ & $0.3840\pm 0.0250$ & $-0.4110\pm 0.1230$  \\
ACT+Planck-cut+DESI+DESY5 & $66.93\pm 0.55$ & $0.3620\pm 0.0230$ & $-0.2832\pm 0.1140$  \\
SPT+Planck-cut+DESI+DESY5 & $66.76\pm 0.52$ & $0.3620\pm 0.0220$ & $-0.2860\pm 0.1120$  \\

\hline
$\bm{\textbf{I}\Lambda\textbf{CDM4}}$ &  &  &    \\
ACT+WMAP+DESI+DESY5 & $67.88\pm 0.58$ & $0.3120\pm 0.0120$ & $-0.0280\pm 0.0380$  \\
SPT+WMAP+DESI+DESY5 & $67.28\pm 0.61$ & $0.3250\pm 0.0140$ & $-0.0840\pm 0.0450$  \\
ACT+Planck-cut+DESI+DESY5 & $67.74\pm 0.53$ & $0.3130\pm 0.0110$ & $-0.0320\pm 0.0310$  \\
SPT+Planck-cut+DESI+DESY5 & $67.74\pm 0.53$ & $0.3090\pm 0.0110$ & $-0.0170\pm 0.0330$  \\
\hline
\hline
\end{tabular}
\end{table*}

In the final step, we analyze the results derived from the SPT CMB constraints. In Figure~\ref{fig1}, constraints in the $\Omega_{\mathrm{m}}$--$\beta$ plane are displayed, with the left panel including the SPT+DESI+DESY5 data (black dashed line) and the right panel including those from SPT+DESI+PantheonPlus (brown dashed line). When using the SPT+DESI+DESY5, the constraint values of $\beta$ are $-0.2310\pm 0.0760$ (I$\Lambda$CDM1), $0.0013\pm 0.0026$ (I$\Lambda$CDM2), $-0.4170\pm 0.1220$ (I$\Lambda$CDM3), and $-0.1080\pm 0.0490$ (I$\Lambda$CDM4). In this case, we observe a slight decrease in overall constraining power due to the significantly larger error bars associated with the SPT data and the absence of information on the lensing power spectrum. For example, for I$\Lambda$CDM1, SPT+DESI+DESY5 gives $\sigma(H_0) = 0.58\,\rm{km\,s^{-1}\,Mpc^{-1}}$, $\sigma(\Omega_{\rm m}) = 0.0210$, and $\sigma(\beta) = 0.0760$, which are 9.4\%, 16.7\%, and 18.8\% worse, respectively, compared to those from Planck+DESI+DESY5. The deviations of $\beta$ from zero are $3.0\sigma$ (I$\Lambda$CDM1), $0.5\sigma$ (I$\Lambda$CDM2), $3.4\sigma$ (I$\Lambda$CDM3), $2.2\sigma$ (I$\Lambda$CDM4). It is worth noting that the enhanced evidence for interaction in I$\Lambda$CDM3 should be mainly interpreted as a dataset-dependent shift of the central value of $\beta$, rather than as a consequence of a tighter constraint, as can be seen from Table~\ref{tab2}. This may be related to the relatively weaker constraining power and different parameter preference of the SPT data in this extended model. Recent studies have also shown that cosmological constraints can depend non-negligibly on the choice of CMB datasets, including ACT and SPT data \cite{Giare:2024oil}. We find that, for the I$\Lambda$CDM2 model, the deviation decreases to $0.5\sigma$ relative to the value obtained using Planck and ACT, which suggests that no interaction between DE and DM. Conversely, the more negative values of $\beta$ in the I$\Lambda$CDM1 and I$\Lambda$CDM3 models indicate stronger support for an interaction in which DE decays into DM. In this case, $\beta$ is anti-correlated with ${\Omega_{\rm m}}$, as clearly seen in Figure~\ref{fig1}, leading to a higher central value of ${\Omega_{\rm m}}$. In particular, when using the SPT+DESI+DESY5 data for the I$\Lambda$CDM1 and I$\Lambda$CDM3 models, the constraint values of ${\Omega_{\rm m}}$ are $0.3620\pm 0.0210$ and $0.3840\pm 0.0250$, respectively. When using the SPT+DESI+PantheonPlus data, the constraint values of $\beta$ are $-0.1460\pm 0.0820$ (I$\Lambda$CDM1), $0.0020\pm 0.0026$ (I$\Lambda$CDM2), $-0.2770\pm 0.1330$ (I$\Lambda$CDM3), and $-0.0580\pm 0.0510$ (I$\Lambda$CDM4). Replacing DESY5 with PantheonPlus weakens the evidence for interaction, yet the deviation of $\beta$ from 0 still reaches $1.8\sigma$ and $2.1\sigma$ for the I$\Lambda$CDM1 and I$\Lambda$CDM3 models, respectively. Therefore, this also results in a corresponding decrease in the values of ${\Omega_{\rm m}}$. For example, the I$\Lambda$CDM1 and I$\Lambda$CDM3 models give the constraint values of ${\Omega_{\rm m}}$ as $0.3390\pm 0.0220$ and $0.3530\pm 0.0260$, respectively. It is worth noting that the current DESI preference for dynamical DE is, to some extent, a manifestation of the $\Omega_{\mathrm{m}}$ tension, with detailed discussions on the $\Omega_{\mathrm{m}}$ inconsistencies available in refs.~\cite{Colgain:2024mtg,Tang:2024lmo}. Interestingly, in the I$\Lambda$CDM1 and I$\Lambda$CDM3 models, the interaction where DE decays into DM leads to an increase in $\Omega_{\mathrm{m}}$, which also results in a slight inconsistency with the $\Omega_{\mathrm{m}}$ values in the $\Lambda$CDM model.

Overall, we find that different CMB datasets influence the evidence for the existence of interaction. In most cases, SPT provides more significant evidence for interaction compared to Planck and ACT. Furthermore, the evidence for existence of interaction is weakened by PantheonPlus, and this result is unaffected by the use of different CMB data. 

To gain a better understanding of the impact of interactions on the expansion history and the role of DESI data, we compare the best-fit predictions of $D_\mathrm{M}(z)$ and $H(z)$ for the $\Lambda$CDM, I$\Lambda$CDM2, and I$\Lambda$CDM3 models using the SPT+DESI+DESY5 data, as shown in Figure~\ref{fig2}. Note that in the right panels showing $H(z)$, both the DESI data points and the theoretical predictions have been rescaled using a fiducial sound horizon of $r_{\rm d} = 147.78\,\mathrm{Mpc}$, corresponding to the Planck best-fit value~\cite{Planck:2018vyg}, for visual comparison only. We observe that the theoretical distance predictions of the $\Lambda$CDM and I$\Lambda$CDM2 models are largely consistent, as the value of $\beta$ in the I$\Lambda$CDM2 model is very small (indicating a very weak interaction), suggesting that the I$\Lambda$CDM2 model is largely compatible with the null hypothesis, i.e., the $\Lambda$CDM model. Conversely, the I$\Lambda$CDM3 model exhibits interactions, leading to some deviations from the theoretical distance predictions of the $\Lambda$CDM model. Interestingly, I$\Lambda$CDM3 explains the values of $D_\mathrm{M}(z)/(r_{\rm d}\sqrt{z})$ at $z = 0.706$, $0.936$, and $1.321$ more successfully than $\Lambda$CDM, with no significant deviations between the other BAO measurements and the best-fit predictions of I$\Lambda$CDM3. Furthermore, as seen in the theoretical predictions for $H(z)$, the DESI data points at $z = 0.510$ and $z = 0.706$ remain inadequately explained by the $\Lambda$CDM, I$\Lambda$CDM2, and I$\Lambda$CDM3 models. However, I$\Lambda$CDM3 provides a better fit to $H(z)$ at $z = 1.321$, simultaneously predicting a smaller present-day expansion rate $H_0$, which results from the decay of DE into DM.

\subsection{Joint constraints from ACT, SPT, WMAP, and Planck}\label{sec3.2}

\begin{figure*}[htpb]
\includegraphics[scale=0.7]{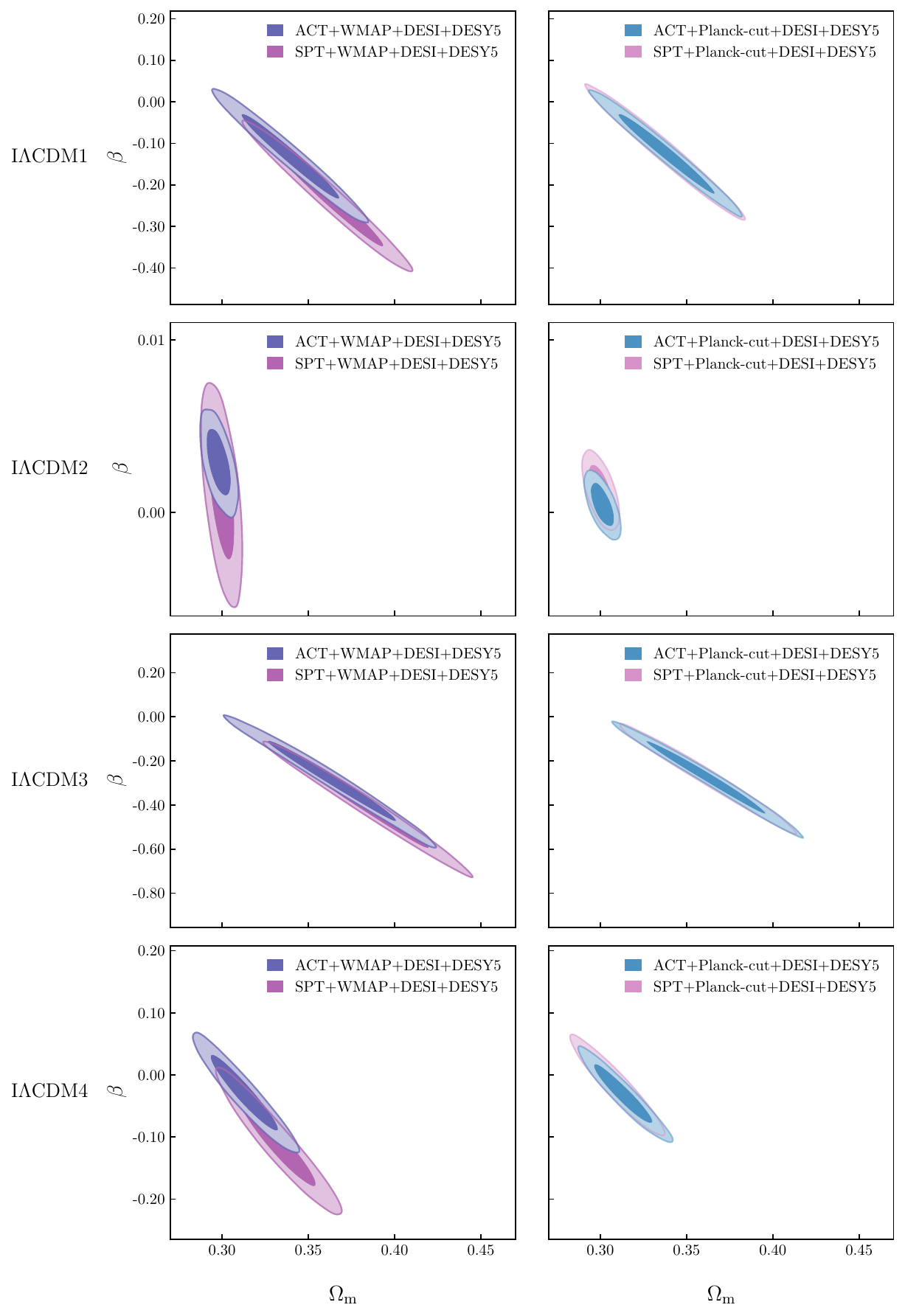}
\centering
\caption{\label{fig3} Two-dimensional marginalized contours ($1\sigma$ and $2\sigma$ confidence levels) in the $\Omega_{\mathrm{m}}$--$\beta$ plane by using the ACT+WMAP+DESI+DESY5, SPT+WMAP+DESI+DESY5, ACT+Planck-cut+DESI+DESY5, and SPT+Planck-cut+DESI+DESY5 data in the I$\Lambda$CDM1, I$\Lambda$CDM2, I$\Lambda$CDM3, and I$\Lambda$CDM4 models.}
\end{figure*}

The constraints on cosmological parameters from CMB dataset combinations are summarized in Table~\ref{tab3}, and the two-dimensional posterior contours in the $\Omega_{\mathrm{m}}$--$\beta$ plane are shown in Figure~\ref{fig3}.

We begin by combining the large-scale temperature and polarization measurements from WMAP with small-scale measurements from either ACT or SPT. In this procedure, cross-covariances between WMAP and ACT, as well as between WMAP and SPT, are disregarded; the datasets are combined only at the likelihood level. A brief clarification is warranted: given the overlap in multipole coverage among these experiments, one might be concerned about potential correlations between WMAP and the ACT and SPT experiments. However, both the ACT and SPT collaborations have explicitly noted that any such correlations can be safely neglected. ACT and SPT primarily probe small angular scales (high $\ell$), whereas WMAP mainly covers large angular scales (low $\ell$). Consequently, the overlap is limited and not consequential; in the few multipole bands where overlap does occur, WMAP’s uncertainties are sufficiently large that its effective weight is subdominant to that of ACT and SPT. A foreseeable outcome is that incorporating WMAP’s large-scale temperature and polarization measurements will not materially affect the constraints on the coupling parameter $\beta$. WMAP provides information around the first acoustic peak and helps break degeneracies among cosmological parameters-most notably between the scalar spectral index $n_{\rm s}$ and the baryon energy density $\Omega_{\rm b} h^2$ (see also refs.~\cite{ACT:2020gnv, AtacamaCosmologyTelescope:2025blo}).

\begin{figure*}[htpb!]
\includegraphics[scale=0.5]{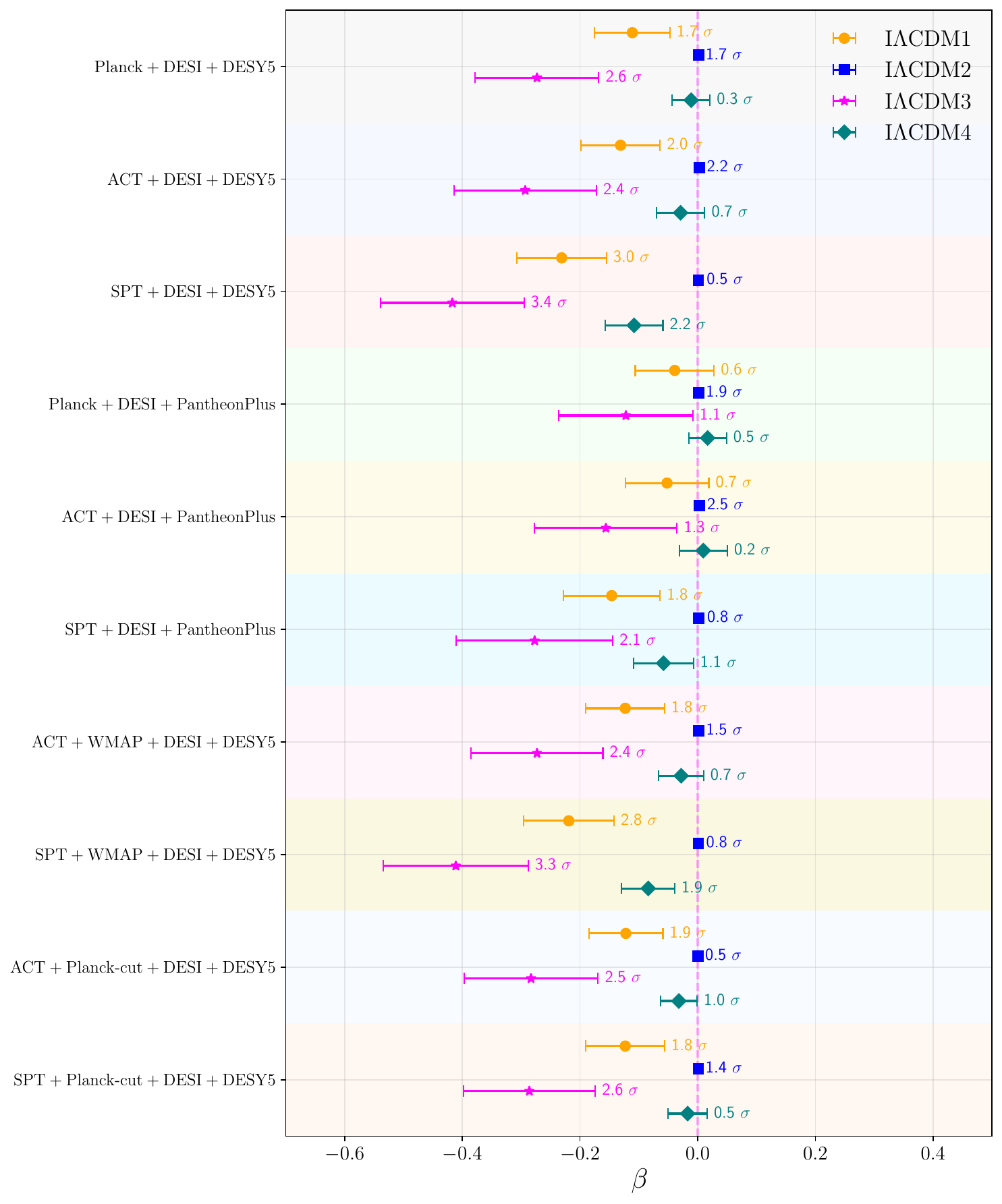}
\centering
\caption{\label{fig4} Whisker plot of $\beta$ in the four I$\Lambda$CDM models, with the $1\sigma$ confidence level constraints for the various datasets employed in our analysis. Additionally, the figure also shows the deviation of $\beta$ relative to 0.}
\end{figure*}

In the case of WMAP+ACT+DESI+DESY5, the constraint values of $ \beta $ are $-0.1230\pm 0.0670$ (I$\Lambda$CDM1), $0.0017\pm 0.0011$ (I$\Lambda$CDM2), $-0.2730\pm 0.1120$ (I$\Lambda$CDM3), and $-0.0280\pm 0.0380$ (I$\Lambda$CDM4), while for the WMAP+SPT+DESI+DESY5 dataset, we obtain $ \beta $ values of $-0.2190\pm 0.0770$ (I$\Lambda$CDM1), $0.0013\pm 0.0017$ (I$\Lambda$CDM2), $-0.4110\pm 0.1230$ (I$\Lambda$CDM3), and $-0.0840\pm 0.0450$ (I$\Lambda$CDM4). We find that the joint constraints on the IDE models obtained by combining WMAP, ACT or SPT, DESI, and DESY5 are essentially consistent with the results obtained from using ACT or SPT combined with DESI and DESY5 alone. As mentioned earlier, the precision of the constraints remains largely unchanged, with the central values slightly shifting towards zero due to the inclusion of the large-scale data. For the I$\Lambda$CDM1 and I$\Lambda$CDM3 models, the value of $\beta$ deviates from zero by $2.8\sigma$ and $3.3\sigma$ when using SPT+WMAP+DESI+DESY5, respectively. These results further confirm that, based on current observational data, there is significant evidence for the existence of interactions in the I$\Lambda$CDM3 model.

\begin{table*}[htbp]
\centering
\caption{\label{tab4}Summary of the $\ln \mathcal{B}_{ij}$ (where $i$ = I$\Lambda$CDM; $j$ = $\Lambda$CDM) values quantifying the evidence of these models relative to the $\Lambda$CDM model using current observational datasets. A positive value indicates a preference for the I$\Lambda$CDM model.}
\setlength{\tabcolsep}{6mm}
\renewcommand{\arraystretch}{1.2} 
\resizebox{\textwidth}{!}{%
\footnotesize
\begin{tabular}{lcccc} 
\hline\hline
\textbf{Data} & \textbf{I$\Lambda$CDM1} & \textbf{I$\Lambda$CDM2} & \textbf{I$\Lambda$CDM3} & \textbf{I$\Lambda$CDM4} \\
\hline
Planck+DESI+DESY5 & $0.41$ & $-7.51$ & $1.67$ & $-3.61$ \\
ACT+DESI+DESY5 & $-0.81$ & $-4.37$ & $1.09$ & $-3.23$ \\
SPT+DESI+DESY5 & $1.12$ & $-6.32$ & $2.13$ & $-4.21$ \\
Planck+DESI+PantheonPlus & $-0.33$ & $-3.83$ & $0.64$ & $-1.39$ \\
ACT+DESI+PantheonPlus & $-1.17$ & $-4.32$ & $0.20$ & $-3.28$ \\
SPT+DESI+PantheonPlus & $3.97$ & $-0.63$ & $4.76$ & $2.33$ \\
ACT+WMAP+DESI+DESY5 & $-2.01$ & $-6.02$ & $0.66$ & $-3.69$ \\
SPT+WMAP+DESI+DESY5 & $1.59$ & $-6.65$ & $3.09$ & $-1.24$ \\
ACT+Planck-cut+DESI+DESY5 & $-0.34$ & $-7.49$ & $0.85$ & $-2.28$ \\
SPT+Planck-cut+DESI+DESY5 & $-0.78$ & $-1.18$ & $6.14$ & $-1.54$ \\
\hline\hline
\end{tabular}
}
\end{table*}

In order to obtain a comprehensive and robust result, we further use the large angular scale Planck data to replace WMAP, combining it with the small angular scale data from ACT and SPT. This is because Planck provides more precise and comprehensive information compared to WMAP, particularly in the region of $\ell \lesssim 30$ (especially in the E-mode polarization measurements), where WMAP lacks such information. To perform this test, we consider the following reduced Planck likelihood:
\begin{itemize}
    \item \textbf{Planck-cut:} We keep the low-$\ell$ \texttt{commander} likelihood for the TT spectrum and the low-$\ell$ \texttt{SimAll} likelihood for the EE spectrum (both at $\ell \lesssim 30$); and retain the high-$\ell$ \texttt{Plik} likelihood for the TT, TE, and EE spectra, focusing only on the reduced multipole range $30 \leq \ell \lesssim 1000$.
\end{itemize}

Note that we retain the Planck temperature and polarization measurements at large angular scales while cutting the high-multipole measurements at $\ell = 1000$ to avoid double-counting the same portion of the sky already measured by ACT and SPT data. Our approach is consistent with that of the ACT collaboration; for further details, see refs~\cite{ACT:2020gnv, AtacamaCosmologyTelescope:2025blo}. This reduced information in the Planck dataset allows us to combine this likelihood with the full ACT DR6 temperature, polarization, and lensing likelihoods, as well as with the full SPT temperature and polarization likelihoods.

We observe that the combination ACT+Planck-cut+DESI+DESY5 provides the tightest constraints, with the constraint values in the I$\Lambda$CDM2 model being $H_0 = 68.37 \pm 0.37~\mathrm{km~s^{-1}~Mpc^{-1}}$, $\Omega_{\mathrm{m}} = 0.3011 \pm 0.0043$, and $\beta = 0.0004 \pm 0.0008$. Furthermore, in the I$\Lambda$CDM2 model, $\beta$ is closer to 0, which further demonstrates no interaction between DE and DM. In other IDE models, it is indicated that there is an interaction. For example, the values of $\beta$ in I$\Lambda$CDM1 and I$\Lambda$CDM3 are $-0.1220\pm 0.0630$ and $-0.2832\pm 0.1140$, indicating deviations from 0 of $1.9\sigma$ and $2.5\sigma$, respectively. Compared to the same data combinations where WMAP is replaced by Planck-cut, the preference for interaction is weakened, which may be due to the fact that WMAP data have lower precision and lack E-mode information at $\ell \lesssim 30$. Moreover, we find that, when using SPT+Planck-cut+DESI+DESY5, the constraint results for the IDE models are essentially consistent with those from ACT+Planck-cut+DESI+DESY5 (with only slight differences), as clearly shown in the right panel of Figure~\ref{fig3}.

\begin{table*}[htbp]
\centering
\caption{The $\chi^{2}_{\rm MAP}$ contributions from CMB, BAO, and SN data for the $\Lambda$CDM, I$\Lambda$CDM2, and I$\Lambda$CDM3 models.}
\label{tab:chi2}
\setlength{\tabcolsep}{6mm}
\renewcommand{\arraystretch}{1.2}
\small
\begin{tabular}{l c c c c}
\hline 
\hline
Model/Data & $\chi^2_{\rm MAP, \ CMB}$ & $\chi^2_{\rm MAP, \ BAO}$ & $\chi^2_{\rm MAP, \ SN}$ & $\chi^2_{\rm MAP, \ total}$ \\
\hline
$\bm{\textbf{$\Lambda$CDM}}$ &  &  &  & \\
Planck+DESI+DESY5 & $10976.25$ & $12.86$ & $1649.02$ & $12638.13$  \\
ACT+DESI+DESY5 & $5904.91$ & $13.79$ & $1649.21$ & $7567.92$  \\
SPT+DESI+DESY5 & $1362.58$ & $12.70$ & $1648.90$ & $3024.19$  \\
\hline
$\bm{\textbf{I$\Lambda$CDM2}}$ &  &  &  & \\
Planck+DESI+DESY5 & $10977.36$ & $11.92$ & $1650.05$ & $12639.34$  \\
ACT+DESI+DESY5 & $5905.64$ & $12.32$ & $1650.89$ & $7568.85$  \\
SPT+DESI+DESY5 & $1363.52$ & $11.57$ & $1650.87$ & $3025.96$  \\
\hline
$\bm{\textbf{I$\Lambda$CDM3}}$ &  &  &  & \\
Planck+DESI+DESY5 & $10976.65$ & $14.96$ & $1640.67$ & $12632.28$  \\
ACT+DESI+DESY5 & $5910.05$ & $10.96$ & $1638.60$ & $7559.62$  \\
SPT+DESI+DESY5 & $1362.89$ & $9.64$ & $1638.97$ & $3011.51$  \\

\hline
\hline
\end{tabular}
\end{table*}

To more intuitively represent the deviation of the coupling parameters for the IDE models across different data combinations, we present a Whisker plot of $\beta$, with the $1\sigma$ confidence level constraints, as shown in Figure~\ref{fig4}. We find that, in all cases, the deviation of $\beta$ from zero is most significant for I$\Lambda$CDM3, with the significance ranging from approximately $1.1\sigma$ to $3.4\sigma$. Specifically, when SPT data are included, both the SPT+DESI+DESY5 and SPT+WMAP+DESI+DESY5 dataset combinations provide support for non-vanishing interactions at approximately the $3.4\sigma$ level, with the marginalized constraints deviating significantly from the $\Lambda$CDM limit. Although replacing DESY5 with PantheonPlus data weakens the significance of the interaction, the SPT+DESI+PantheonPlus still provide evidence for interaction at the $2.1\sigma$ level. Similarly, combining Planck-cut data also weakens the significance of the interaction, but the SPT+Planck-cut+DESI+DESY5 combination still provides evidence for interaction at the $2.6\sigma$ level. Additionally, for I$\Lambda$CDM1, in most cases, the deviation of $\beta$ from zero is observed at the $0.6\sigma \sim 3.0\sigma$ level, with SPT+DESI+DESY5 providing $3.0\sigma$ evidence supporting the presence of interactions. Notably, we observe from the plot that both I$\Lambda$CDM2 and I$\Lambda$CDM4 also exhibit deviations in the range of $0.2\sigma \sim 2.5\sigma$, but their $\beta$ central values are very small, primarily centered around zero. This is mainly due to the well-constrained errors, as discussed earlier. Specifically, for the I$\Lambda$CDM2 model, for example, ACT+DESI+PantheonPlus gives $\beta = 0.0032\pm 0.0013$ (with a deviation of $2.5\sigma$ from zero). Therefore, we conclude that the interaction strength for I$\Lambda$CDM2 is very weak. 

Overall, in most data combinations (especially those including SPT and DESY5), the IDE models that follow $Q \propto \rho_{\rm de}$ are more likely to support the existence of interaction between DE and CDM compared to the $Q \propto \rho_{\rm c}$ model, especially the $Q = \beta H_0 \rho_{\rm de}$ (I$\Lambda$CDM3) model.

\subsection{Bayesian evidence for model comparison}\label{sec3.3}

We use the Bayesian evidence selection criterion to choose the preferred I$\Lambda$CDM models over the $\Lambda$CDM model based on the current observational data. To compute the Bayesian evidence of the models, we employ the publicly available code {\tt MCEvidence}\footnote{\url{https://github.com/yabebalFantaye/MCEvidence}.} \cite{Heavens:2017afc,Heavens:2017hkr}. The Bayesian evidence $Z$ is defined as
\begin{equation}
Z = \int_{\Omega} P(D|\bm{\theta},M)P(\bm{\theta}|M)P(M)\ {\rm d}\bm{\theta},
\label{eq: lnZ}
\end{equation}
where $P(D|\bm{\theta},M)$ is the likelihood of the observational data $D$ given the parameters $\bm{\theta}$ and the cosmological model $M$, $P(\bm{\theta}|M)$ is the prior probability of $\bm{\theta}$ given $M$, and $P(M)$ is the prior of $M$. We calculate the Bayes factor in logarithmic space, defined as $\ln \mathcal{B}_{ij} = \ln Z_i - \ln Z_j$, where $Z_i$ and $Z_j$ represent Bayesian evidence of two models.

The strength of model preference is typically assessed using the Jeffreys scale \cite{Kass:1995loi,Trotta:2008qt}.  According to this scale: if $\left|\ln \mathcal{B}_{ij}\right|<1$, the evidence is inconclusive; $1\le\left|\ln \mathcal{B}_{ij}\right|<2.5$ represents weak evidence; $2.5\le\left|\ln \mathcal{B}_{ij}\right|<5$ is moderate; $5\le\left|\ln \mathcal{B}_{ij}\right|<10$ is strong; and if $\left|\ln \mathcal{B}_{ij}\right|\ge 10$, the evidence is decisive. 

In Table~\ref{tab4}, we show the Bayes factors $\ln \mathcal{B}_{ij}$ for the I$\Lambda$CDM models relative to the $\Lambda$CDM model, based on the current observational data. Here, $i$ denotes the I$\Lambda$CDM model, and $j$ denotes the $\Lambda$CDM model. It is important to emphasize that a positive value indicates a preference for the IDE models over the $\Lambda$CDM model, while negative values indicate a preference for the $\Lambda$CDM model. We find that the Bayes factor values for I$\Lambda$CDM3 are positive in all cases of data combinations. Specifically, for all cases including SPT data, $\ln \mathcal{B}_{ij} = 2.13$, $4.76$, $3.09$, and $6.14$ for SPT+DESI+DESY5, SPT+DESI+PantheonPlus, SPT+WMAP+DESI+DESY5, and SPT+Planck-cut+DESI+DESY5, respectively, indicating moderate to strong favor towards the I$\Lambda$CDM3 model relative to the $\Lambda$CDM model. By contrast, for I$\Lambda$CDM2, the Bayes factor values are negative in all cases of data combinations. For example, the values of $\ln \mathcal{B}_{ij} = -7.51$, $-6.32$, $-6.02$, $-6.65$, and $-7.49$ for Planck+DESI+DESY5, SPT+DESI+DESY5, ACT+WMAP+DESI+DESY5, SPT+WMAP+DESI+DESY5, and ACT+Planck-cut+DESI+DESY5, respectively, indicate strong evidence disfavoring I$\Lambda$CDM2 relative to the $\Lambda$CDM model. For I$\Lambda$CDM1, some cases provide weak to moderate evidence in favor relative to $\Lambda$CDM, while others show comparable favor to the $\Lambda$CDM model. For I$\Lambda$CDM4, only SPT+DESI+PantheonPlus gives $\ln \mathcal{B}_{ij} = 2.33$, indicating weak preference relative to $\Lambda$CDM. However, in all other cases, there is weak to moderate evidence in favor of the $\Lambda$CDM model. Interestingly, in some cases, a high statistical significance for the interaction parameter (e.g., the $\sim 3.4\sigma$ deviation of $\beta$) does not necessarily result in a Bayes factor in favor of the extended model. This apparent discrepancy arises because the two quantities evaluate the models from different statistical perspectives. The marginalized parameter posterior is a local measure of parameter estimation, indicating that the posterior distribution favors a nonzero value of $\beta$ within the extended model. In contrast, the Bayes factor is a global measure of model selection. As an integral over the full multidimensional prior parameter space, the Bayes factor intrinsically penalizes the extended model for its additional free parameters, which is not strictly correlated with the local significance of a single parameter.

Based on the results of the Bayesian analysis, and to better understand the contribution of each data component, we take the Planck+DESI+DESY5, ACT+DESI+DESY5, and SPT+DESI+DESY5 combinations as examples. We calculated the individual minimum $\chi^2_\mathrm{MAP}$ values for the $\Lambda$CDM, I$\Lambda$CDM2 (the least favored model), and I$\Lambda$CDM3 (the most favored model), which are summarized in Table~\ref{tab:chi2}. We find that the IDE models yield a slightly worse fit (i.e., higher $\chi^2_\mathrm{MAP}$) to the CMB data compared to $\Lambda$CDM, particularly for the I$\Lambda$CDM3 model, which features stronger interactions. This occurs because, in the I$\Lambda$CDM3 model, DE decays into DM, leading to an inferred increase in DM density. This enhancement tends to raise the amplitude of the CMB acoustic peaks, resulting in a slight degradation of the fit to the CMB data. However, the lower total $\chi^2_\mathrm{MAP}$ value of the I$\Lambda$CDM3 model compared to the $\Lambda$CDM model is primarily driven by the contribution of late-time data, particularly the SN data. Overall, the current observational data favor these IDE models with $Q \propto \rho_{\rm de}$ relative to those with $Q \propto \rho_{\rm c}$ compared to $\Lambda$CDM, especially the $Q = \beta H_0 \rho_{\rm de}$ (I$\Lambda$CDM3) model.

\section{Conclusion}\label{sec4}

The DESI collaboration has utilized DR1 or DR2 BAO measurements, and when combined with CMB data and DESY5 SN data, these have revealed a $\sim 4\sigma$ deviation from the $\Lambda$CDM paradigm, suggesting a preference for dynamical DE \cite{DESI:2024mwx,DESI:2025zgx}. Given the present cosmological preference for dynamical DE, it is natural to ask whether this deviation from $\Lambda$CDM could instead be accounted for by a non-zero interaction between DE and DM. In our previous study, we used DESI DR1 data combined with Planck CMB data and DESY5 SN data, and reported a $3\sigma$ preference for the IDE model with $Q = \beta H_0 \rho_{\rm de}$ \cite{Li:2024qso}. Recently, some studies indicate that considering different SN data and various CMB experiment data can significantly impact this dynamical DE preference \cite{Giare:2024oil,Huang:2024qno}. Therefore, in this context, it is crucial to examine whether the interaction between DE and DM is influenced by the latest observational data, in order to provide comprehensive and robust results. 

In this work, our aim is to conduct a comprehensive analysis using the latest observational data to determine whether there is an interaction between DE and DM and to identify which IDE model is more supported. We use the BAO data from DESI DR2, CMB data from Planck, WMAP, ACT, and SPT, as well as SN data from DESY5 and PantheonPlus to constrain the IDE models. We consider four phenomenological IDE models, i.e., I$\Lambda$CDM1 ($Q=\beta H\rho_{\rm de}$), I$\Lambda$CDM2 ($Q=\beta H\rho_{\rm c}$), I$\Lambda$CDM3 ($Q=\beta H_0\rho_{\rm de}$), and I$\Lambda$CDM4 ($Q=\beta H_0\rho_{\rm c}$). We explore the extent to which CMB experiments other than Planck support an interaction between DE and DM, and test whether replacing DESY5 with PantheonPlus reduces the preference for such an interaction.

Our overall analysis indicates that, the IDE models following $Q \propto \rho_{\rm de}$ are more likely to support the existence of interaction compared to those following $Q \propto \rho_{\rm c}$. In most cases, the evidence for a non-zero interaction is strongest in the I$\Lambda$CDM3 model, with I$\Lambda$CDM1 ranking second. In contrast, I$\Lambda$CDM2 and I$\Lambda$CDM4 yield $\beta$ values that are very small, indicating at most a very weak interaction. We find that, in most cases, CMB experiments other than Planck tend to strengthen the evidence for the preference of interaction between DE and DM. For example, in the I$\Lambda$CDM3 model, SPT+DESI+DESY5 yields $\beta = -0.4170 \pm 0.1220$, with a deviation from zero of $3.4\sigma$, which significantly strengthens the evidence for interaction, compared to Planck+DESI+DESY5, which gives $\beta = -0.2730 \pm 0.1050$ at the $2.6\sigma$ level. Additionally, regardless of the CMB experiment data considered, replacing DESY5 with PantheonPlus SN data results in a reduced evidence for interaction. For example, SPT+DESI+PantheonPlus gives $\beta = -0.2770\pm 0.1330$, indicating $2.1\sigma$ evidence supporting the presence of interaction. Furthermore, when we combine ACT or SPT with WMAP, we find that the results for the interaction are essentially unaffected, with the constraints from the former still being slightly stronger than those from the latter. However, when combining ACT or SPT with Planck-cut, the evidence for interaction weakens, with both providing comparable constraints. Specifically, ACT+Planck-cut+DESI+DESY5 and SPT+Planck-cut+DESI+DESY5 yield values of $\beta = -0.2832 \pm 0.1140$ and $\beta = -0.2860 \pm 0.1120$ in the I$\Lambda$CDM3 model, providing about $2.5\sigma$ evidence supporting the presence of interaction.  

We evaluate the Bayesian evidence for IDE models relative to $\Lambda$CDM using current observational data. Our analysis shows that the Bayes factors for I$\Lambda$CDM3 are positive across all dataset combinations, and are positive in most combinations for I$\Lambda$CDM1, indicating a weak-to-strong preference for these models over $\Lambda$CDM. Conversely, across all dataset combinations, the Bayes factor values for I$\Lambda$CDM2 are uniformly negative, indicating weak-to-strong evidence in favor of $\Lambda$CDM; I$\Lambda$CDM4 follows the same pattern, except for the SPT+DESI+PantheonPlus case. Overall, the current observational data favor these IDE models with $Q \propto \rho_{\rm de}$ relative to those with $Q \propto \rho_{\rm c}$, compared to $\Lambda$CDM. 

In summary, these findings warrant caution when assessing the robustness of the preference for an interaction between DE and DM inferred from DESI BAO, SN, and CMB data. Further tests and forthcoming data, including CMB-S4 measurements \cite{CMB-S4:2016ple}, more precise late-time observations from Large Synoptic Survey Telescope \cite{LSST:2008ijt} and Euclid \cite{Euclid:2024yrr}, and subsequent DESI BAO releases, will be essential to reach a more definitive conclusion.

\section*{Acknowledgments}
We thank Hao Wang, Sheng-Han Zhou, Yan-Hong Yao, and Peng-Ju Wu for their helpful discussions. We are also grateful to William Giar\`e, Ze-Yu Peng, and Ye-Huang Pang for their helpful correspondence. This work was supported by the National Natural Science Foundation of China (Grants Nos. 12533001, 12575049, 12473001, and 12473091), the National SKA Program of China (Grants Nos. 2022SKA0110200 and 2022SKA0110203), the China Manned Space Program (Grant No. CMS-CSST-2025-A02), the National 111 Project (Grant No. B16009), and the Fundamental Research Funds for the Central Universities (Grant No. N2405008).


\bibliography{main}

\end{document}